\documentclass{PoS}

\title{Status of the (p)MSSM Higgs sector}

\ShortTitle{Status of the (p)MSSM Higgs sector}

\author{\speaker{A.~Arbey}$^,$\thanks{Also Institut Universitaire de France, 103 boulevard Saint-Michel, 75005 Paris, France.}$^{~,1,2}$, M.~Battaglia$^{2,3}$, A.~Djouadi$^{4}$, F.~Mahmoudi$^{\dagger,1,2}$, M.~M\"uhlleitner$^5$, G.~Robbins$^{1,6,7}$, M.~Spira$^8$\\
\\
        $^1$Univ. Lyon, Univ. Lyon 1, CNRS/IN2P3, Institut de Physique Nucl\'eaire de Lyon,\\ UMR5822, F-69622 Villeurbanne, France\vspace*{0.2cm}\\
        $^2$CERN, 1 Esplanade des Particules, CH-1211 Geneva 23, Switzerland\vspace*{0.2cm}\\
        $^3$University of California at Santa Cruz, Santa Cruz Institute of Particle Physics, CA 95064, USA\vspace*{0.2cm}\\
        $^4$Universit\'e Grenoble Alpes, USMB, CNRS, LAPTh, F-74000 Annecy, France\vspace*{0.2cm}\\
        $^5$Institute for Theoretical Physics, Karlsruhe Institute of Technology, 76128 Karlsruhe, Germany\vspace*{0.2cm}\\
        $^6$National Institute of Chemical Physics \& Biophysics, R\"avala 10, 10143 Tallinn, Estonia\vspace*{0.2cm}\\
        $^7$Univ. Lyon, Univ. Lyon 1, ENS de Lyon, CNRS, Centre de Recherche Astrophysique de Lyon UMR5574, F-69230 Saint-Genis-Laval, France\vspace*{0.2cm}\\
        $^8$Paul Scherrer Institut, CH-5232 Villigen PSI, Switzerland\\
        \\
        E-mails: \email{alexandre.arbey@ens-lyon.fr}, \email{marco.battaglia@ucsc.edu}, \email{abdelhak.djouadi@cern.ch}, \email{nazila@cern.ch}, \email{milada.muehlleitner@kit.edu}, \email{glenn.robbins@kbfi.ee}, \email{michael.spira@psi.ch}}

\abstract{
We present some highlights on the complementaries of the Higgs and SUSY searches at the LHC, using the 8 and 13 TeV results. In particular, we discuss the constraints that can be obtained on the MSSM parameters by the determination of the Higgs boson mass and couplings.
In addition, we investigate the interplay with heavy Higgs searches, and evaluate how higher LHC luminosities and a future linear collider can help probing the pMSSM Higgs sector and reconstructing the underlying parameters.
}

\FullConference{The 39th International Conference on High Energy Physics (ICHEP2018)\\
		4-11 July, 2018\\
		Seoul, Korea}

\begin{document}

\section{Introduction}

The discovery of the Higgs boson at the LHC has marked a major step for our understanding of particle physics, and for the construction of the Higgs sector of new physics scenarios. Direct searches for new particles are currently actively persued at the LHC, in particular in the context of supersymmetry (SUSY). No new physics signal has been discovered so far, implying that new physics should be subtle or heavy. Therefore, indirect constraints are at the moment of utmost importance. The measurements of the properties of the Higgs boson can provide in this respect very strong constraints on new physics scenarios. The measurement of its mass at 125 GeV \cite{Tanabashi:2018oca} is very constraining for supersymmetry, because the Higgs mass can receive large corrections from the stop sector, and has a large impact on the SUSY parameter space~\cite{Arbey:2011ab}. In the following, we will discuss the status of the Higgs sector of the phenomenological MSSM.
To do so, we perform random scans on the 19 parameters of the pMSSM, following the procedures detailled in~\cite{Arbey:2011un}. In particular, we use a master program based on SuperIso~\cite{Mahmoudi:2008tp}, generate the MSSM spectra with SOFTSUSY~\cite{Allanach:2001kg} and compute the Higgs boson decay widths and couplings with HDECAY~\cite{Djouadi:1997yw}. We keep only the parameter points where the lightest supersymmetric particle is a neutralino (constituting a dark matter candidate) and a light Higgs mass of $125\pm3$ GeV.

\section{Higgs coupling measurements and SUSY direct searches}

\begin{figure}[t!]
\begin{center}
\hspace*{-0.5cm}\includegraphics[width=.35\textwidth]{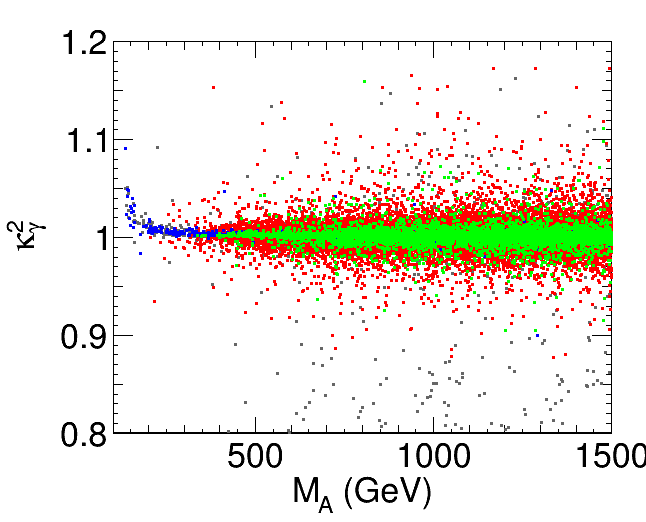}\includegraphics[width=.35\textwidth]{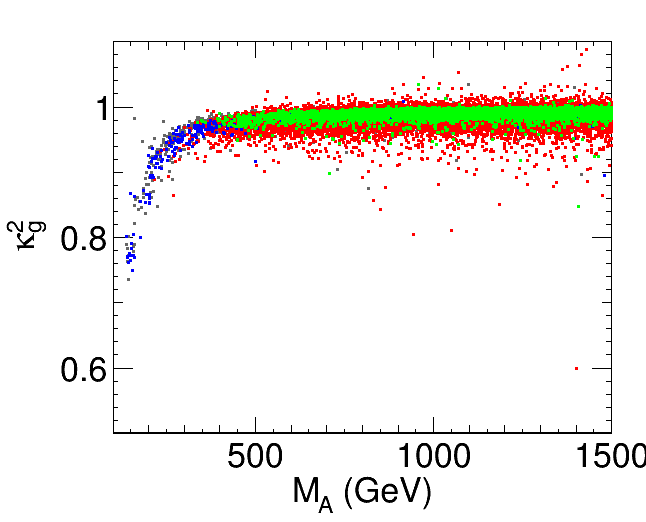}\includegraphics[width=.35\textwidth]{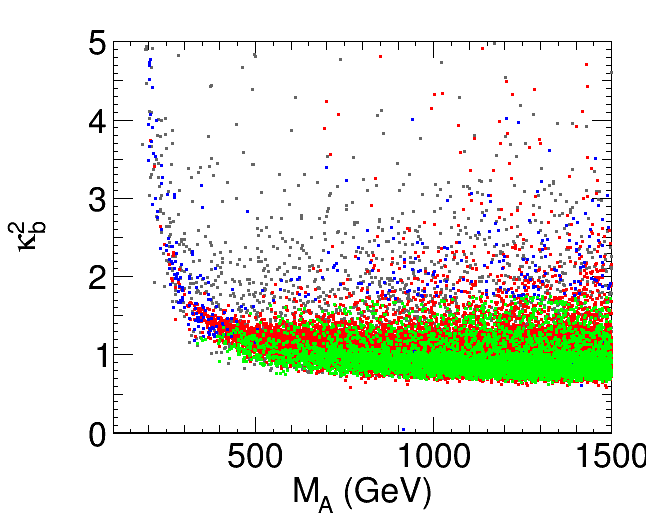}\vspace*{-0.5cm}
\end{center}
\caption{Distributions of the squared light scalar Higgs couplings to photons (left), gluons (center) and bottoms (right), as a function of $M_{A}$ in the pMSSM. The grey points correspond to all points with $M_h\sim125$ GeV, the red ones pass in addition the LEP constraints, the blue points are also consistent with LHC SUSY direct searches and the green points are compatible with Higgs coupling measurements.\label{figcoups}}
\end{figure}

We first study the interplay of the measurement of the Higgs boson properties and of the results of the SUSY direct searches. We impose the LEP constraints on the SUSY masses \cite{Tanabashi:2018oca}. To assess the constraints from SUSY searches at the LHC, we generate events with PYTHIA \cite{Sjostrand:2014zea}, simulate the detector with Delphes \cite{deFavereau:2013fsa} and obtain constraints from ATLAS and CMS results with 36 fb$^{-1}$~\cite{Ventura:2017itv} for gluino and squark, neutralino and chargino, stop and sbottom, and monojet searches. For the Higgs measurements, we consider that there are 6 independent effective Higgs couplings, to the photons $\kappa_\gamma$, gluons $\kappa_g$, vector bosons $\kappa_V$, tops $\kappa_t$, bottoms $\kappa_b$ and taus $\kappa_\tau$. We combined the ATLAS and CMS measurements of the Higgs couplings at 7+8 TeV \cite{Khachatryan:2016vau} and 13 TeV \cite{Brandstetter:2018eju}, and to check if a point is consistent with these measurements, we use a $\chi^2$ test and keep only points in agreement at 95\% C.L. In Figure \ref{figcoups}, we present the photon, gluon and bottom squared coupling distributions as a function of $M_{A}$, applying different sets of constraints. All the shown couplings are sensitive to $M_A$, in addition to other SUSY parameters which modify the couplings at loop level. In particular, the photon and gluon couplings are sensitive to the stop and sbottom masses. The bottom coupling is modified by the $\Delta_b$ corrections \cite{Djouadi:2005gj}. We see that the combination of the direct searches and Higgs measurements strongly restricts the coupling values to be close to 1. Since the different couplings are related to SUSY masses, these results can be used to obtain constraints on the pMSSM parameters.

\section{Heavy Higgs direct searches and Higgs coupling measurements}

\begin{figure}[h!]
\begin{center}
\hspace*{-0.5cm}\includegraphics[width=.35\textwidth]{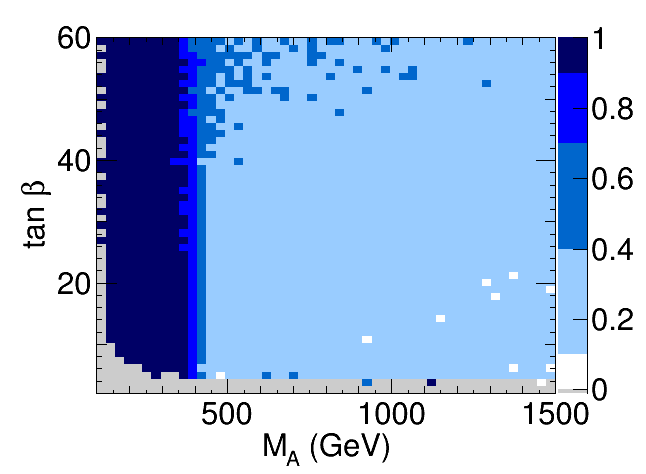}\includegraphics[width=.35\textwidth]{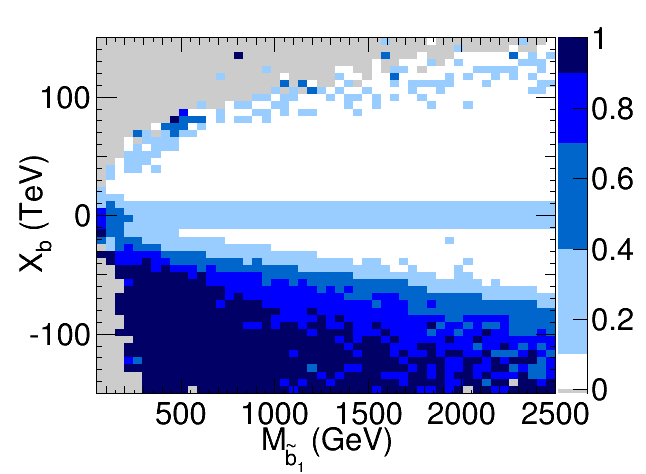}\includegraphics[width=.35\textwidth]{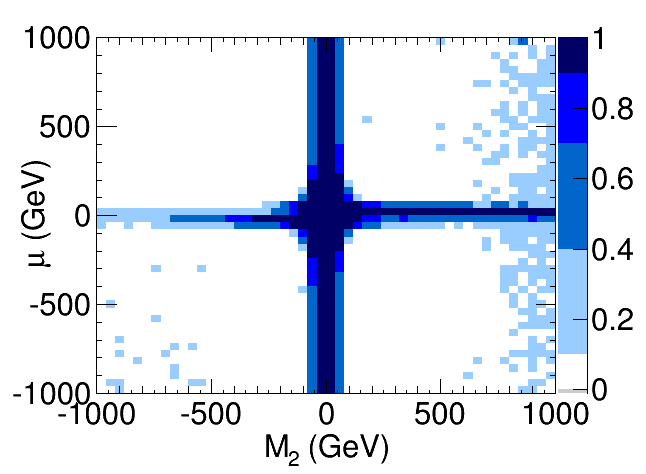}\\
\hspace*{-0.5cm}\includegraphics[width=.35\textwidth]{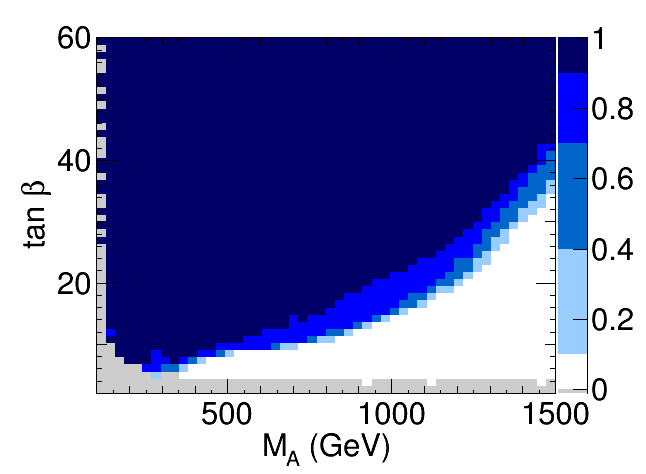}\includegraphics[width=.35\textwidth]{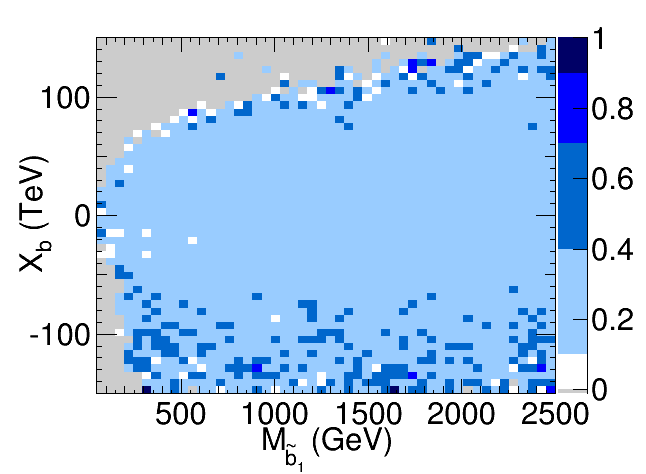}\includegraphics[width=.35\textwidth]{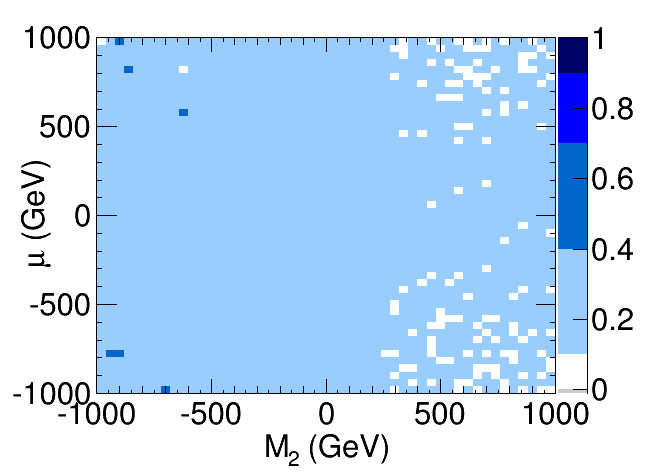}
\caption{Fraction of excluded points by Higgs coupling measurements (top) and heavy Higgs searches (bottom), in the $(M_A,\tan\beta)$ (left), $(M_{\tilde{b}_1},X_b=A_b-\mu\times\tan\beta)$ (center) and $(M_2,\mu)$ (right) parameter planes.\label{figheavy}}
\end{center}
\end{figure}

Another way to constrain the Higgs sector is through searches for heavier Higgs states. We use HDECAY~\cite{Djouadi:1997yw} and SusHi~\cite{Harlander:2012pb} to compute the heavy Higgs decay rates and production cross-sections, respectively, and apply the ATLAS and CMS heavy Higgs search limits \cite{Schaarschmidt:2018ask}. We compare the exclusion from the Higgs coupling measurements to the one from heavy Higgs searches in Figure~\ref{figheavy}, which reveals the important interplay between the light Higgs coupling measurements and the heavy Higgs search limits: While $(M_A,\tan\beta)$ is very strongly constrained by $H/A \to \tau^+\tau^-$ searches, the $(M_{\tilde{b}_1},X_b)$ and $(M_2,\mu)$ parameter planes are more constrained by the Higgs coupling measurements.

\section{Prospects for the MSSM Higgs sector}

\begin{table}[t!]
\begin{center}

\begin{tabular}{|c|cc|cc|cc|}
  \hline
  & $M_A ({\rm GeV})$ & $\tan\beta$ & $M_A ({\rm GeV})$ & $\tan\beta$ & $M_A ({\rm GeV})$ & $\tan\beta$ \\ \hline
 Original parameters & 334.9 & 9.9 & 427.3 & 5.7 & 657.2 & 12.7\\
HL-LHC recontruction & 394$\pm$40 & 9.6$\pm$4.0 & 471$^{+341}_{-56}$ & - & - & -\\
ILC recontruction & 351$\pm$23 & 9.2$\pm$1.9 & 460$^{+54}_{-45}$ & 10.4$^{+6}_{-4}$ & 747.7$^{+302}_{-97}$ & 10.2$^{+20}_{-4}$\\
\hline
\end{tabular}~\\[0.1cm]
\begin{tabular}{|c|c|c|c|c|c|}
\hline
Original $\mu \tan\beta~({\rm TeV})$ &  $-149.9$ & $-86.6$  & 0 &  79.6 & 108.6 \\ \hline
 ILC recontruction & $-76.3^{+28}_{-39}$ & $-124.6^{+46}_{-60}$ & $-2.2\pm22$ &  67.2$^{+39}_{-22}$& 82.5$^{+40}_{-22}$ \\
\hline
\end{tabular}
\caption{Reconstruction potential of different pMSSM scenarios with HL-LHC and ILC projections.\label{tabfuture}}
\end{center}
\end{table}

We now study the prospects for the high-luminosity LHC (HL-LHC) run and ILC \cite{Moortgat-Picka:2015yla}, by considering the possibility to reconstruct specific scenarios using the Higgs coupling measurements. We test two categories of scenarios: the first one where only $M_A$ and $\tan\beta$ are varied, and the second where $\mu \tan\beta$ is modified. We assume the accuracy reached when the ILC collects 1 ab$^{-1}$ of luminosity at energies between 350 and 800 GeV. Table~\ref{tabfuture} summarises our results for several example scenarios (some at the limit of being excluded by current searches). We can conclude that the HL-LHC alone would allow us to reconstruct CP-odd Higgs masses up to 500 GeV. For higher masses, or for scenarios with modified $\mu \tan\beta$, the ILC will be necessary to identify the underlying parameters of the scenario.

\end{document}